\documentclass[12pt]{article}
\usepackage{a4wide,amssymb,float}
\usepackage{latexsym,epsfig,graphicx}

\def\be{\begin{equation}}
 \def\ee{\end{equation}}
 \def\bea{\begin{eqnarray}}
 \def\eea{\end{eqnarray}}

 \def\o{\omega}
 
\newcommand{\fr}{\frac}

\def\2{\frac{1}{2}}
\def\4{\frac{1}{4}}

\catcode`\@=11

\@addtoreset{equation}{section}

\def\@normalsize{\@setsize\normalsize{15pt}\xiipt\@xiipt
\abovedisplayskip 14pt plus3pt minus3pt%
\belowdisplayskip \abovedisplayskip
\abovedisplayshortskip  \z@ plus3pt%
\belowdisplayshortskip  7pt plus3.5pt minus0pt}
\def\small{\@setsize\small{13.6pt}\xipt\@xipt
\abovedisplayskip 13pt plus3pt minus3pt%
\belowdisplayskip \abovedisplayskip
\abovedisplayshortskip  \z@ plus3pt%
\belowdisplayshortskip  7pt plus3.5pt minus0pt
\def\@listi{\parsep 4.5pt plus 2pt minus 1pt
            \itemsep \parsep
            \topsep 9pt plus 3pt minus 3pt}}

\def\underline#1{\relax\ifmmode\@@underline#1\else
        $\@@underline{\hbox{#1}}$\relax\fi}
\@twosidetrue \relax

\catcode`@=12

\catcode`\@=11

\def\section{\@startsection{section}{1}{\z@}{3.5ex plus 1ex minus
   .2ex}{2.3ex plus .2ex}{\large\bf}}

\def\ps@headings{\def\@oddfoot{}\def\@evenfoot{}
\def\@oddhead{\hbox{}\hfill
        \makebox[.5\textwidth]{\raggedright\ignorespaces --\thepage{}--
        \hfill }}
\def\@evenhead{\@oddhead}
\def\subsectionmark##1{\markboth{##1}{}}
}

\ps@headings

\catcode`\@=12

\begin{document}

\begin{titlepage}
\begin{flushright}

\end{flushright}

\vspace{0.30in}
\begin{centering}

{\large {\bf Discontinuities in Scalar Perturbations of
Topological Black Holes}}
\\

\vspace{0.7in} {\bf George Koutsoumbas,$^{a,*}$\,\,\,\,
Eleftherios
Papantonopoulos$^{a,\flat}$\,\,\,\,\\ \vspace{0.14in} and George
Siopsis$^{b,\natural}$}

\vspace{0.44in}

$^{a}$Department of Physics, National Technical University of Athens,\\
Zografou Campus GR 157 73, Athens, Greece\\
$^{b}$Department of Physics and Astronomy, The University of
Tennessee, Knoxville, TN 37996 - 1200, USA

\vspace{0.04in}

\vspace{0.04in}

\end{centering}

\vspace{0.8in}

\begin{abstract}

We study the perturbative behaviour of topological black holes. We
calculate both analytically and numerically the quasi-normal modes
of scalar perturbations. In the case of small black holes we find
discontinuities of the quasi-normal modes spectrum at the critical
temperature and we argue that this is evidence of a second-order
phase transition.

 \end{abstract}

\begin{flushleft}

\vspace{0.9in} $^{*}$~kutsubas@central.ntua.gr \\
$^{\flat}$~lpapa@central.ntua.gr\\
$^\natural$~siopsis@tennessee.edu

\end{flushleft}
\end{titlepage}

\section{Introduction}

The knowledge of the spectrum of the quasi-normal modes (QNMs) is
a powerful tool in the study of the late time behaviour of black
holes. The QNMs are the complex frequencies by which a black hole
responds if it is initially perturbed, and they do not depend on
the details of the initial perturbation but rather on the
intrinsic features of the black hole itself. The radiation
associated with these modes is expected to be seen with
gravitational wave detectors in the coming years, giving valuable
information on the properties of black holes.

The QNMs of black holes in asymptotically flat spacetimes have
been extensively studied and their spectrum was computed
numerically and in many cases also analytically (for reviews,
see~\cite{KS,N}). The advances in string theory and mainly the
Anti-de Sitter - conformal field theory (AdS/CFT) correspondence
has renewed the interest of computing the QNMs of black holes in
asymptotically AdS
spacetimes~\cite{CM,HH,CL,WLA,kokkotas,Konoplya}. Recent results
from the Relativistic Heavy Ion Collider \cite{RHIC} show that a
thermal quark-gluon plasma (QGP) is formed which is strongly
coupled. The AdS/CFT correspondence provides the connection
between the QGP and string black holes. In \cite{Friess:2006kw}
the quasi-normal modes of AdS$_{5}$-Schwarzschild black hole have
been calculated and it was shown that they provide a dual
description of the fluctuations of the QGP. Recently, in a more
phenomenological approach, the AdS/CFT correspondence was applied
to condensed matter physics~\cite{condensed_Matter}. It was
shown~\cite{Hartnoll:2008vx} that fluctuations of the metric and
the background electromagnetic field determines the conductivity
of the boundary theory.

According to the AdS/CFT correspondence, a large static black hole
in AdS corresponds to an (approximately) equilibrium thermal state
in the CFT~\cite{cft}. In ref.~\cite{HH} it was shown that the
QNMs for the scalar perturbations of large Schwarzschild-AdS black
holes scaled with the temperature and it was argued that the
perturbed system in the dual description will approach to thermal
equilibrium of the boundary conformal field theory. However, when
the black hole size is comparable to the AdS length scale there is
a clear departure from this behaviour. It was then conjectured
that this behaviour may be connected with a Hawking-Page phase
transition \cite{phase1, phase2} which occurs when the temperature
lowers.

For small black holes the behaviour of QNMs is not very well
understood. For the case of Schwarzschild-AdS black holes the QNMs
do not scale linearly with the temperature any more and their
connection with the boundary conformal field theory is not clear.
In~\cite{Koutsoumbas:2006xj,Koutsoumbas:2008pw} we studied the
QNMs of electromagnetic and gravitational perturbations of
topological black holes (TBHs) in AdS spacetime coupled to a
scalar or an electromagnetic field. For small black holes,
compared to the length scale  of the AdS space, we found that the
QNMs behave quite differently from the QNMs of large black holes.
Near the critical temperature purely dissipative modes appeared in
the spectrum and the QNMs change slope around that point. We
attributed this behaviour to a second order phase transition of
the charged TBH towards the AdS vacuum solution.

In the literature there is a discussion of possible connections
between the classical and thermodynamical properties of black
holes~\cite{Reall:2001ag}. In particular the question whether the
knowledge of the QNM spectrum can give information about
thermodynamical phase transitions in a wider class of black holes
has recently gained considerable interest. It was suggested
in~\cite{Jing:2008an} that the Dirac and Rarita-Schwinger
perturbations are related to thermodynamic phase transitions of
charged black holes. This has been criticised
in~\cite{Berti:2008xu}, the argument being that the relation
between the QNMs and the phase transition had not been properly
formulated. In this direction, the discontinuities observed in the
heat capacity of charged Kaluza-Klein black holes with squashed
horizons, were connected with the quasi-normal
spectrum~\cite{He:2008im}. Further evidence of a non-trivial
relation between the thermodynamical and dynamical properties of
black holes was provided in~\cite{Shen:2007xk}.

In this work we calculate both analytically and numerically the
QNMs of scalar perturbations of topological-AdS black holes in
$d=4,5$ and $6$ dimensions. In all dimensionalities considered, we
find a discontinuity of the QNMs spectrum at the critical point.
This provides further evidence of a second order phase transition
at the critical point observed
in~\cite{Koutsoumbas:2006xj,Koutsoumbas:2008pw}. In our numerical
calculations we used the method developed in~\cite{HH}. We
conjecture that this method is problematic for large $n$ because
it breaks down, and a regularization scheme is proposed.

Our work is organized  as follows: in section 2 the analytical
calculation is presented, section 3 contains comments on the
numerical method used, and in section 4 the numerical results may
be found. Finally the conclusions are presented in section 5.

\section{Analytical Calculation}
\label{sec2}

Analytical results for the QNMs of topological black holes are
only known for the massless case. In four dimensions they were
calculated in~\cite{Aros:2002te} while a generalization to
d-dimensions have been presented in \cite{Birmingham:2006zx}. For
the general case early numerical results have been discussed in
\cite{CM} and in \cite{Wang:2001tk}. To study the phase transition
in topological black holes the QNM spectrum has to be known, at
least near the critical point. Therefore in this section we
calculate analytically the QNMs of scalar perturbations of
topological black holes in various dimensions. At the critical
point the mass goes to zero so we reproduce the known exact
results. Also we discuss the asymptotic form of QNMs for large
black holes. In five dimensions we present analytical results
expressible through the Heun function. The analytical results will
give us a clear indication of a second order phase transition
around the critical point which will be substantiated by numerical
results presented in the next section.

We consider the bulk action \be I=\frac{1}{16\pi G} \int
d^{d}x\sqrt{-g}\left[ R+\frac{(d-1)(d-2)}{l^{2}}\right]~,\ee in
asymptotically AdS$_{d}$ where $G$ is the Newton's constant and
$l$ is the AdS radius. The presence of
 a negative cosmological constant $(\Lambda=-\frac{(d-1)(d-2)}{2l^{2}})$
  allows the existence of black holes with a
topology $\mathbb{R} \times \Sigma$, where $\Sigma $ is a
$(d-2)$-dimensional manifold of constant negative curvature. These
black holes are known as topological black holes. The simplest
solution of this kind reads
\begin{equation}
ds^2 =
-f(r) dt^2 + \frac{dr^2}{f(r)} + r^2 d\sigma^2  \ , \ \ f(r) = r^2
- 1 - \frac{2\mu}{r^{d-3}}~,  \label{Top-BH-Einstein}
\end{equation}
where we employed units in which the AdS radius is $l=1$ and
$d\sigma$ is the line element of $\Sigma $. The latter is locally
isomorphic to the hyperbolic manifold $H^{d-2}$ and of the form
\be \Sigma =H^{d-2}/\Gamma \quad\quad , \quad \Gamma \subset
O(d-2,1)\;, \ee where $\Gamma$ is a freely acting discrete
subgroup (i.e., without fixed points) of isometries. {The geometry
of the topological black holes  as well their basic properties
have been studied extensively in the
literature~\cite{Lemos}-\cite{Myung:2006tg}.

The configurations (\ref{Top-BH-Einstein}) are asymptotically
locally AdS spacetimes. It has been  shown in
\cite{Gibbons:2002pq} that the massless configurations where
$\Sigma $ has negative constant curvature are stable under
gravitational perturbations. More recently the stability of the
TBHs was discussed in \cite{Birmingham:2007yv}.

We shall calculate analytically the quasi-normal modes of scalar
perturbations  of these black holes following the method discussed
in \cite{Koutsoumbas:2008pw}. A numerical approach will be
discussed in section \ref{sec3}.

Massless scalar perturbations
obey the radial wave equation \be \frac{1}{r^{d-2}} ( r^{d-2} f(r)
\Phi')' + \frac{\omega^2}{f(r)} \Phi - \frac{\xi^2 + (
\frac{d-3}{2} )^2}{r^2} \Phi = 0~. \ee By defining \be \Phi =
r^{-\frac{d-2}{2}} \Psi \ee the wave equation can be cast into a
Schr\"odinger-like form, \be\label{sch}
  -\frac{d^2\Psi}{dr_*^2}+V[r(r_*)]\Psi =\o^2\Psi \;,
\ee in terms of the tortoise coordinate defined by
\be\label{tortoise}
  \frac{dr_*}{dr} = \frac{1}{f(r)}\,,
\ee and the potential is given by \be\label{eqVV} V(r) = f(r)
\left\{ \frac{d(d-2)}{4} + \frac{k^2-\frac{3}{4}}{r^2} +
\frac{(d-2)^2\mu}{2r^{d-1}} \right\} \ \ , \ \ \ \ k^2 = \xi^2 +
\Big{(}\frac{d-3}{2}\Big{)}^2~. \ee For later convenience we
define  $\Lambda=k^2-3/4$.


To obtain analytic expressions for the quasi-normal frequencies,
it is convenient to introduce the coordinate
\be\label{eqru} u = \left( \frac{r_+}{r} \right)^2~. \ee The wave
equation (\ref{sch}) becomes \be\label{eq13} 4 u^{3/2} \hat f(u)
(u^{3/2} \hat f(u) \Psi')' + [ \hat\omega^2 - \hat V ] \Psi = 0 \
\ , \ \ \ \ \hat\omega = \frac{\omega}{r_+}~,\ee where prime
denotes differentiation with respect to $u$ and we have defined
\bea\label{eq14} \hat f(u) \equiv \frac{f(r)}{r_+^2} &=&
\frac{1}{u} - \frac{1}{r_+^2} - \frac{2\mu}{r_+^{d-1}}
u^{\frac{d-3}{2}}~, \nonumber \\ \hat V(u) \equiv
\frac{V(r)}{r_+^2}&=& \hat f(u) \left\{ \frac{d(d-2)}{4} +
\hat\Lambda u + \frac{(d-2)^2\mu}{2r_+^{d-1}} u^{\frac{d-1}{2}}
\right\}~, \eea and \be \hat\Lambda = \frac{\Lambda}{r_+^2}\ \ , \
\ \ \ 2\mu = r_+^{d-3} (r_+^2 - 1)~. \ee

\subsection{Exact solution in $d=5$}

In five dimensions ($d=5$) we may solve the wave equation
in terms of a Heun function and use the latter to determine the spectrum exactly albeit numerically.
By factoring $\hat f(u)$,
\[ \hat f(u) = (1-u)\left( \frac{1}{u} +1 - \frac{1}{r_+^2} \right) \]
we may obtain an exact solution to the wave
equation in terms of a Heun function,
\be \Psi (u) =
u^{-3/4} \left( u + \frac{1}{1-1/r_+^2} \right)^{-\frac{\hat\omega
\sqrt{1-1/r_+^2}}{2(2-1/r_+^2)}}
(1-u)^{-\frac{i\hat\omega}{2(2-1/r_+^2)}}
\ \mathrm{Heun} (a,q,\alpha,\beta,\gamma,\delta,
1-u) \ee The Heun function obeys the equation
\be z(z-1)(z-a)F'' -
[(-\alpha-\beta-1)z^2+((\delta+\gamma)a-\delta+\alpha+\beta+1)z
-a\gamma]
F' - [-\alpha\beta z+q] F = 0 \ee and the various constants are
\[ a = \frac{2-\frac{1}{r_+^2}}{1-\frac{1}{r_+^2}} \ \ , \ \ \ \
q =\frac{1}{2(1-\frac{1}{r_+^2})(2 -\frac{1}{r_+^2})^2}
 \Big{[}
\frac{1}{2r_+^2} (\xi^2 +1) (2-\frac{1}{r_+^2})^2 + i\hat\omega
(2-\frac{1}{r_+^2}) \]
\[ - \hat\omega^2 (2- \frac{3}{2r_+^2} )  +
\hat\omega (i\hat\omega
-2+\frac{1}{r_+^2})(1-\frac{1}{r_+^2})^{3/2} \Big{]}~, \]
\[ \alpha = \beta = -\frac{(i+\sqrt{1-\frac{1}{r_+^2}})\hat\omega}{2(2
-\frac{1}{r_+^2})} \ \ , \ \ \ \
\gamma = 1- \frac{i\hat\omega}{2-\frac{1}{r_+^2}}
\ \ , \ \ \ \ \delta = -1~.
\]
It behaves nicely at the horizon ($u\to 1$). Requiring $\Psi (0) =
0$ yields the constraint \be\label{eq54} \mathrm{Heun}
(a,q,\alpha,\beta,\gamma,\delta, 1) = 0~, \ee which
may be solved for $\hat\omega$ to obtain the quasi-normal
frequencies of scalar modes in five dimensions.

\subsection{Large Black Holes}

For large black holes, we let $\hat\omega, \hat\Lambda \to 0$,
$r_+\to\infty$. The wave equation becomes \be\label{eq13l} u^{3/2}
[(u^{1/2} - u^{d/2}) \Psi']' - \frac{(d-2) [d + (d-2)
u^{\frac{d-1}{2}}]}{16} \Psi = 0~. \ee Two linearly independent
solutions are \be \Psi = u^{-\frac{d-2}{4}} \ \ , \ \ \ \
u^{-\frac{d-2}{4}} \ln \left( 1- u^{\frac{d-1}{2}} \right)~. \ee
Both are unacceptable as they diverge at $u=0,1$, respectively. It
follows that the only possible frequencies are those which diverge
as $r_+\to\infty$.

In five dimensions, the lowest mode may be deduced from the exact constraint in the limit $r_+\to\infty$ keeping $\xi$ and $\hat\omega$ fixed (and therefore $\hat\Lambda\to 0$, $\omega \sim r_+$).
In this limit,
\[ a = 2 \ \ , \ \ \ \
q = \frac{[2(i-1)
+(i-2)\hat\omega
]\hat\omega}{8}~~,~~~~ \alpha = \beta = -\frac{(i+1)\hat\omega}{4} \ \ , \ \ \ \
\gamma = 1- \frac{i\hat\omega}{2}
\ \ , \ \ \ \ \delta = -1~.
\]
The lowest frequency in five dimensions is found to be
\be \hat\omega = 3.12 - 2.75i \ee
in agreement with earlier numerical results \cite{HH}. Therefore, for a large black hole, $\omega \sim r_+$ and QNMs do not contribute to the macroscopic (hydrodynamic) behaviour of the gauge theory fluid on the AdS boundary.

\subsection{Near the Critical Point}

At the critical point ($r_+=1$, $\mu = 0$), the wave equation
reduces to \be 4u^{1/2} (u^{1/2} (1 -u)\Psi')' + \left[
\frac{\omega^2}{1-u} - \frac{d(d-2)}{4u} - \Lambda \right] \Psi =
0~. \ee The solution which is well-behaved at the boundary is \be
\Psi(u) = (1-u)^{-i\omega/2} u^{d/4} F(
\frac{\frac{d+1}{2}+i\xi-i\omega}{2}, \frac{\frac{d+1}{2} -i\xi
-i\omega}{2} ; \frac{d+1}{2} ; u)~. \ee At the horizon, $u\to 1$,
it behaves as \be \Psi(u) \sim \mathcal{A}_+ (1-u)^{-i\omega/2} +
\mathcal{A}_- (1-u)^{i\omega/2}~~,~~~~ \mathcal{A}_\pm =
\frac{\Gamma(\frac{d+1}{2})\Gamma(\pm i\omega)}{\Gamma(\frac{
\frac{d+1}{2}-i\xi\pm
i\omega}{2})\Gamma(\frac{\frac{d+1}{2}+i\xi\pm i\omega}{2})}~. \ee
Demanding $\mathcal{A}_- = 0$, we deduce the quasi-normal
frequencies \be\label{eqana1} \omega_n = \pm \xi - i \left( 2n +
\frac{d+1}{2} \right) \ \ , \ \ \ \ n = 0,1, 2, \dots \ee which
have finite real part (except in the special case $\xi = 0$). This
result agrees with the quasi-normal frequencies of massless
topological black holes which they were calculated analytically
for the first time in~\cite{Aros:2002te}.

The hypergeometric functions corresponding to the eigenfunctions
are polynomials. Explicitly, \bea \Psi_0 (u)
&=& A_0 (1-u)^{-i\xi/2-(d+1)/4} u^{d/4}~, \nonumber\\
\Psi_1(u) &=& A_1 (1-u)^{-i\xi/2-(d+5)/4} u^{d/4} \left[ 1 +
\frac{2(1+i\xi)}{d+1} u \right]~, \eea etc. They are orthogonal
under the inner product (no complex conjugation!) \be \langle n |m
\rangle \equiv \int_0^1 \frac{du}{u^{1/2} (1-u)} \Psi_n (u) \Psi_m
(u) \ee defined by appropriate analytic continuation of the
parameter $\xi$. To normalize them ($\langle n |n\rangle = 1$),
choose \be A_0^2 = \frac{\Gamma(-i\xi)}{\Gamma (-i\xi -
\frac{d+1}{2})\Gamma (\frac{d+1}{2})} \ \ , \ \ \ \ A_1^2 =
\frac{(d+1)\Gamma(-1-i\xi)}{(-2i\xi -d-3)\Gamma(-i\xi -
\frac{d+5}{2}) \Gamma(\frac{d+1}{2})}~,\ee etc.

Moving away from the critical point, the frequencies shift by \be
\delta\omega_n = \frac{1}{2\omega_n} \frac{\langle n |\mathcal{H}'
|n\rangle}{\langle n|n\rangle}~, \ee where \be \mathcal{H}'
\Psi_n = r_+^2 \left[ -4u^{3/2}\hat
f(u)\left(u^{3/2}\hat f(u)\Psi_n'\right)' + \hat V (u) \Psi_n \right]
-\omega_n^2\Psi_n~, \ee and we applied standard first-order
perturbation theory.

We obtain \bea\label{eqar1} \delta\omega_0 &=& -i\left( r_+ -
1 \right) \frac{2^{d} \Gamma (\frac{d}{2}) \Gamma (-i\xi)}{\sqrt\pi
\Gamma (-i\xi + \frac{d-3}{2} )}~, \nonumber\\
\delta\omega_1 &=& -i\left( r_+ - 1 \right) \frac{2^{d-2} \Gamma
(\frac{d}{2}) \Gamma (-i\xi-1) [-2i\xi (d+1) -d^2+6d-17]}{\sqrt\pi
\Gamma (-i\xi + \frac{d-3}{2} )}~. \eea For small $\xi$, there is
no change in the imaginary part at first order. Below the critical
point ($r_+<1$), the change in the real part is negative
($\delta\omega_n < 0$) and the real part decreases. There are
critical values of $\xi$ (determined by $\xi+\delta\omega_n\approx
0$), \be\label{eqana6a} \xi_0 \approx \sqrt{\frac{2^{d}
\Gamma(\frac{d}{2})}{\sqrt\pi \Gamma( \frac{d-3}{2})}}
\sqrt{1-r_+}~,~~~ \xi_1 \approx \sqrt{\frac{2^{d-2}
\Gamma(\frac{d}{2}) [d^2-6d+17]}{\sqrt\pi \Gamma(
\frac{d-3}{2})}}\sqrt{1-r_+} \ee below which the corresponding
mode does not propagate (purely dissipative mode). It turns out
that for $\xi < \xi_{n}$
 pairs of
purely dissipative modes emerge.

Above the critical point ($r_+ >1$), $\delta\omega_n > 0$ and the
real parts of the modes increase. The modes do not become
purely dissipative for any value of $\xi$.

Also notice that below the critical point, $\delta\omega_n$
increases with $n$, therefore the real part decreases with $n$
(positive slope) whereas above the critical point we obtain a {\em
negative} slope for propagating modes.

These results will be discussed further along with a comparison with numerical results in section \ref{sec4}.

\section{Numerical Calculation}
\label{sec3}

In this section we present the numerical calculation of the QNMs
of scalar perturbations of small topological black holes following the method developed by Horowitz and Hubeny~\cite{HH}.


After performing the transformation
$\Psi(r)=\frac{\psi_\omega(r)}{r^{\frac{d-2}{2}}} e^{-i \omega
r_*},$ the wave equation becomes \be f(r) \fr{d^2\psi_\omega(r)}{d
r^2}+ \left(\fr{d f(r)}{d r} -2 i \omega \right) \fr{d
\psi_\omega(r)}{d r} = V(r) \psi_\omega(r)~.\ee The change of
variables $r=1/x$ yields an equation of the form $$s(x)
\left[(x-x_+)^2 \fr{d^2 \psi_\omega(x)}{d x^2}\right] +t(x)
\left[(x-x_+) \fr{d \psi_\omega(x)}{d x}\right]+ u(x)
\psi_\omega(x)=0~,$$ where $x_+ = 1/r_+$ and $s(x), t(x)$ and
$u(x)$ are  given by
\begin{eqnarray} s(x) &=& \sum_k s_k (x-x_+)^k~,
\nonumber\\ t(x) &=& \sum_k t_k (x-x_+)^k~,\nonumber\\ u(x) &=&
\sum_k u_k (x-x_+)^k~.\nonumber\end{eqnarray} Expanding the wave
function around the (inverse) horizon $x_+$,
\be\psi_\omega(x)=\sum_0^\infty a_n(\omega) (x-x_+)^n~,\ee we
arrive at a recurrence formula for the coefficients, \be
a_n(\omega) = -\fr{1}{P_{n,0}}\sum_{m=n-4}^{n-1} P_{m,n-m}
a_m(\omega)~, \label{reca} \ee \be \ P_{m,n-m} \equiv m (m-1)
s_{n-m} +m t_{n-m} +u_{n-m}~. \label{pnm}\ee We note that the few
coefficients $a_m(\omega)$ with negative index $m$ which will
appear for $n < 2$ should be set to zero, while $a_0(\omega)$ is
set to one. Since the wave function should vanish at infinity $(r
\rightarrow \infty, x=0),$ we deduce \be \psi_\omega(0) \equiv
\sum_0^\infty a_n(\omega) (-x_+)^n=0~.\label{HH} \ee The solutions
of this equation are precisely the quasi-normal frequencies.

The change of variables \be\bar{a}_n \equiv a_n x_+^n\ \ \ \ ,\ \
\bar{P}_{m,n-m} \equiv P_{m,n-m} x_+^{n-m} = m (m-1) \bar{s}_{n-m}
+m \bar{t}_{n-m} +\bar{u}_{n-m}\label{eq34a}\ee transforms eqs.~(\ref{reca}) and (\ref{HH}) into \be \bar{a}_n =
-\fr{1}{\bar{P}_{n,0}}\sum_{m=n-4}^{n-1} \bar{P}_{m,n-m}
\bar{a}_m~,\ee and \be \psi_\omega(0) \equiv \sum_0^\infty
\bar{a}_n(\omega) (-1)^n=0~,\label{HH1} \ee
respectively.

To get a feeling about the solutions of the recurrence relations
we consider the limit of very large $n.$ In this limit we may keep
just the terms of order $n^2$, so relation (\ref{eq34a}) is
approximated by \be \bar{P}_{m,n-m}\simeq m^2\bar{s}_{n-m} \ee and
the recurrence relation reads
$$\bar{a}_n = -\fr{1}{\bar{s}_0}\sum_{m=n-N}^{n-1} \bar{s}_{n-m}
\bar{a}_m~.$$ Notice that it has constant coefficients, so it can
be solved analytically. For simplicity let us restrict attention
to the case $N=4.$ The solution of the recurrence relation reads
$$\bar{a}_n = \frac{K_1}{r_1^n}+ \frac{K_2}{r_2^n}+
\frac{K_3}{r_3^n}+ \frac{K_4}{r_4^n}~.$$ It should be emphasized
that the coefficients $K_m$ ($m=1,2,3,4$) do not depend on $n.$
Explicitly,
$$K_m=\frac{\mathcal{A}_m}{\mathcal{B}_m} \ ,$$
where
\bea \mathcal{A}_m &\equiv&
\bar{s}_0*(\bar{A}_3*r_m^3+\bar{A}_2*r_m^2+\bar{A}_1*r_m+\bar{A}_0)
+\bar{s}_1*(\bar{A}_2*r_m^3+\bar{A}_1*r_m^2+\bar{A}_0*r_m)
\nonumber\\
& & +\bar{s}_2*(\bar{A}_1*r_m^3+\bar{A}_0*r_m^2)
+\bar{s}_3*(\bar{A}_0*r_m^3) \ , \nonumber\\
 \mathcal{B}_m &\equiv&
-(4*\bar{s}_4*r_m^4+3*\bar{s}_3*r_m^3+2*\bar{s}_2*r_m^2+\bar{s}_1)\
. \nonumber\eea $r_m$ are the four roots of the equation \be
\bar{s}_0 + \bar{s}_1 r_m + \bar{s}_2 r_m^2 + \bar{s}_3 r_m^3 +
\bar{s}_4 r_m^4=0~, \label{algeq} \ee respectively and $\bar{A}_k,
\ k=0,1,2,3$ are the values of the first four coefficients which
are necessary to completely specify the solution.


For large $n,$ the sum is dominated by the root which has the
smallest absolute value. If all roots have an absolute value which
is bigger than one, the coefficients will go to zero for large $n$
and the method is expected to work. However, if  the absolute
value of one of the roots is smaller than one, we run into
problems. In our case, one of the roots is equal to $-1.$

To see this, we report expressions for various
coefficients in table \ref{table0}.
\begin{table}[!ht]

\begin{center}
\begin{tabular}{|l|l|l|l|}
\hline $i$ & $\bar s_i$ & $\bar t_i$ & $\bar u_i$ \\ \hline $0$ &
$x_+(x_+^2-3)$ & $-3 x_+ +x_+^3 + 2 i \omega x_+^2$ & $0$ \\
\hline $1$ & $x_+(4 x_+^2-9)$ & $-12 x_+ +6 x_+^3 + 4 i \omega
x_+^2$ & $-3 x_+ +(\frac{3}{4}-\xi^2) x_+^3$ \\ \hline $2$ &
$x_+(6 x_+^2-10)$ & $-18 x_+ +12 x_+^3 + 2 i \omega x_+^2$ & $-3
x_+ +(\frac{5}{2}-2 \xi^2) x_+^3$ \\ \hline $3$ & $x_+(4 x_+^2-5)$
& $-12 x_+ +10 x_+^3$ & $-3 x_+ +(\frac{11}{4}-\xi^2) x_+^3$ \\
\hline $4$ & $x_+(x_+^2-1)$ & $-3 x_+ +3 x_+^3$ & $-x_+ +x_+^3$ \\
\hline
\end{tabular}

\end{center}
\caption{Coefficients used in the numerical method
(eq.~(\ref{eq34a}))}\label{table0}
\end{table}
Using these, one may
show that $$
\bar{P}_{n,0}+\bar{P}_{n-2,2}+\bar{P}_{n-4,4}=\bar{P}_{n-1,1}
+\bar{P}_{n-3,3}~.$$ Notice that this equality holds exactly in
$d=4.$ For $d=5,6$ the two sums differ by an amount which is of order $\mathcal{O} (1)$ so it is
negligible for large $n.$ In particular one may show the relation
$$\bar{s}_{0} +\bar{s}_{2} +\bar{s}_{4} = \bar{s}_{1}
+\bar{s}_{3}~,$$ which implies that $-1$ is a root of eq.~(\ref{algeq}), as advertised.

Thus our case is inconclusive and one should examine
$\frac{1}{n}$ corrections to settle the issue of convergence. We expect that
$|\frac{\bar{a}_{n+1}}{\bar{a}_n}|=1+\frac{\mu}{n}~.$ If $\mu<0$
the sequence will converge to zero and the series can
be finite. Otherwise there is a problem with convergence.

To ensure convergence and obtain numerical solutions of the
equations $\Re[\psi_\omega(0)]=0, \ \ \Im[\psi_\omega(0)]=0~,$ we
shall introduce a regularization scheme as follows: rather than
imposing the condition that the wave function vanish at
$r=\infty,$ we impose the condition that it vanish at
$r=R=\frac{r_+}{\epsilon}, \epsilon <<1.$ This translates to
$x=X=\epsilon x_+.$ Then eq.~(\ref{HH}) is replaced by \be
\psi_\omega(X) \equiv \sum_0^\infty a_n(\omega) (X-x_+)^n=0~
\Rightarrow \sum_0^\infty \bar{a}_n(\omega) (-1)^n
(1-\epsilon)^n=0~. \label{HHa} \ee Asymptotically,
$\bar{a}_n(\omega) = (-1)^n A(\omega)$ which solves the recursion
relation and is confirmed by numerical results. Without the
reqularization parameter ($\epsilon = 0$), this leads to problems
because the left-hand side of (\ref{HH1}) diverges. With a finite
$\epsilon$, the series is asymptotically geometric and converges.

If $\epsilon$ is small enough, the error introduced is negligible.
However, choosing too small a value for $\epsilon$ is not calculationally efficient as it may call for too many
terms in the series (so that the individual terms become small
enough). We have used $\epsilon=0.01$ for $d=4, \ d=5,$ and
$\epsilon=0.02$ for $d=6.$ With these values our results agree
with the analytical results for $r_+ \approx 1.$

\begin{figure}[!t]
\centering
\includegraphics[angle=-90,scale=0.5]{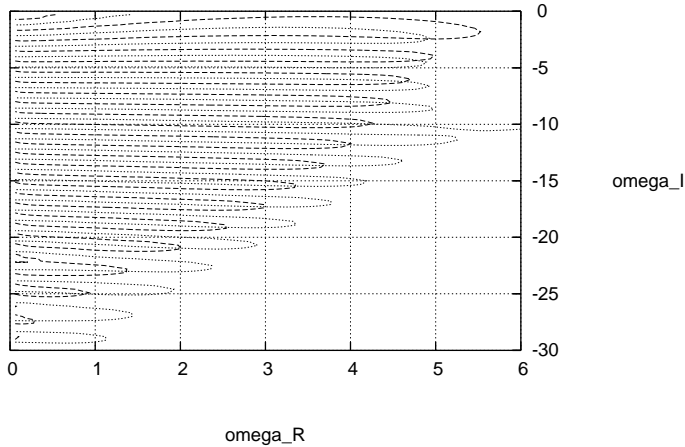}
\caption{Scalar QNMs in four dimensions at $r_+=0.95, \xi=5.0.$}
\label{d4_095_50}
\end{figure}

\begin{figure}[!ht]
\centering
\includegraphics[angle=-90,scale=0.5]{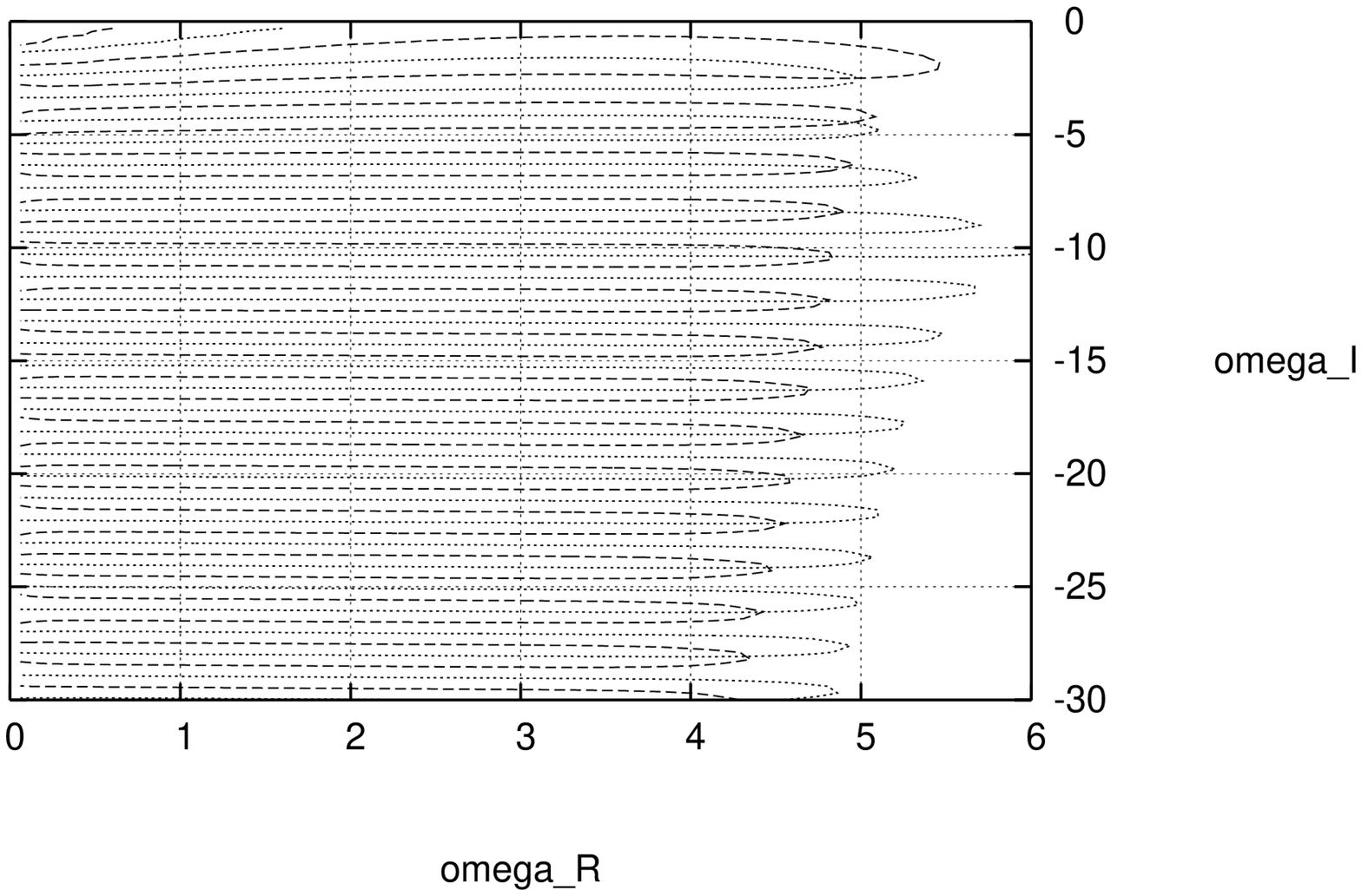}
\caption{Scalar QNMs in four dimensions at $r_+=0.99, \xi=5.0.$}
\label{d4_099_50}
\end{figure}

\section{Results}
\label{sec4}

\subsection{$d=4$}

\begin{figure}[!t]
\centering
\includegraphics[angle=-90,scale=0.5]{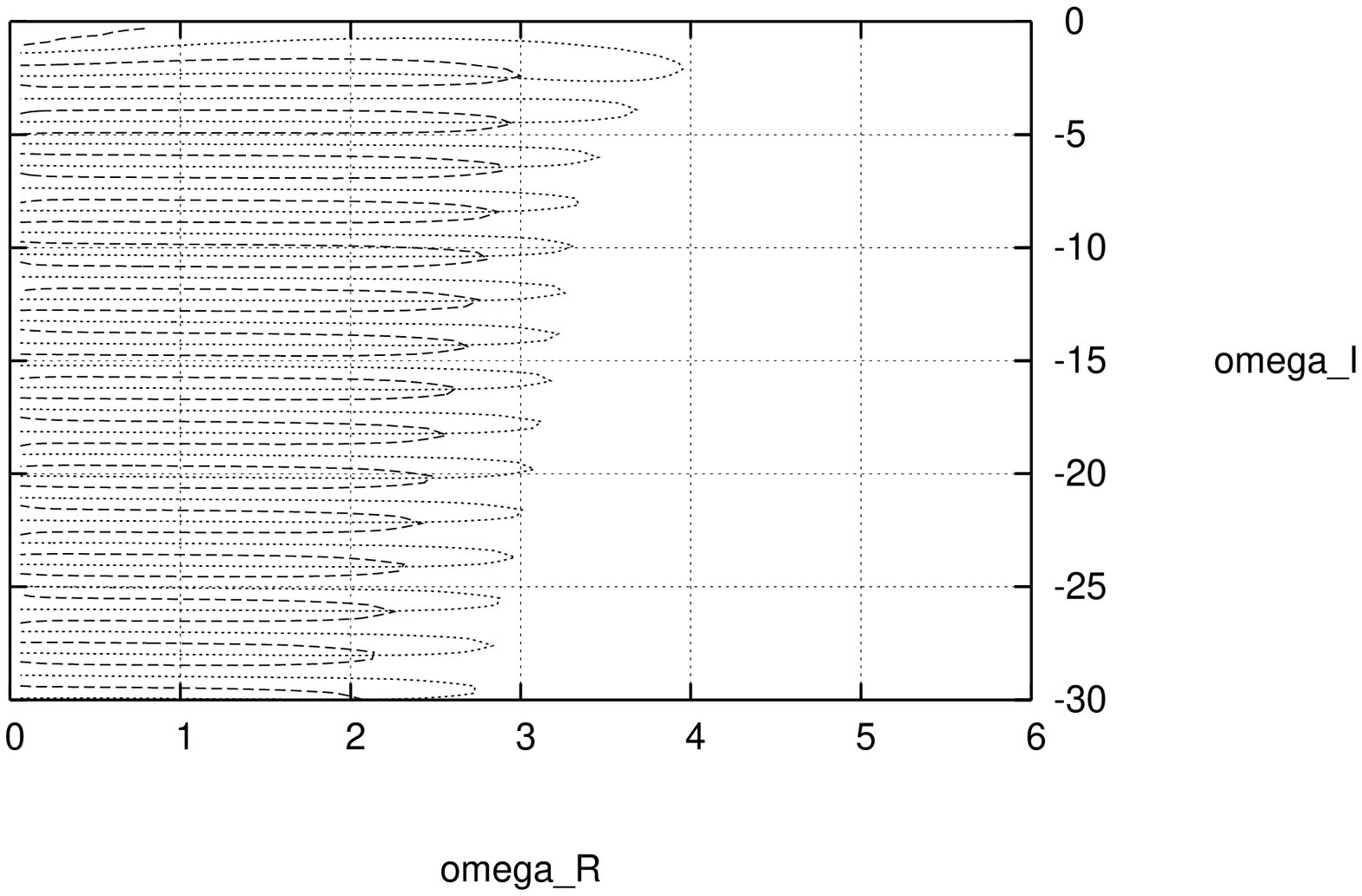}
\caption{Scalar QNMs in four dimensions at $r_+=0.99, \xi=3.0.$}
\label{d4_099_30}
\end{figure}

\begin{figure}[!ht]
\centering
\includegraphics[angle=-90,scale=0.5]{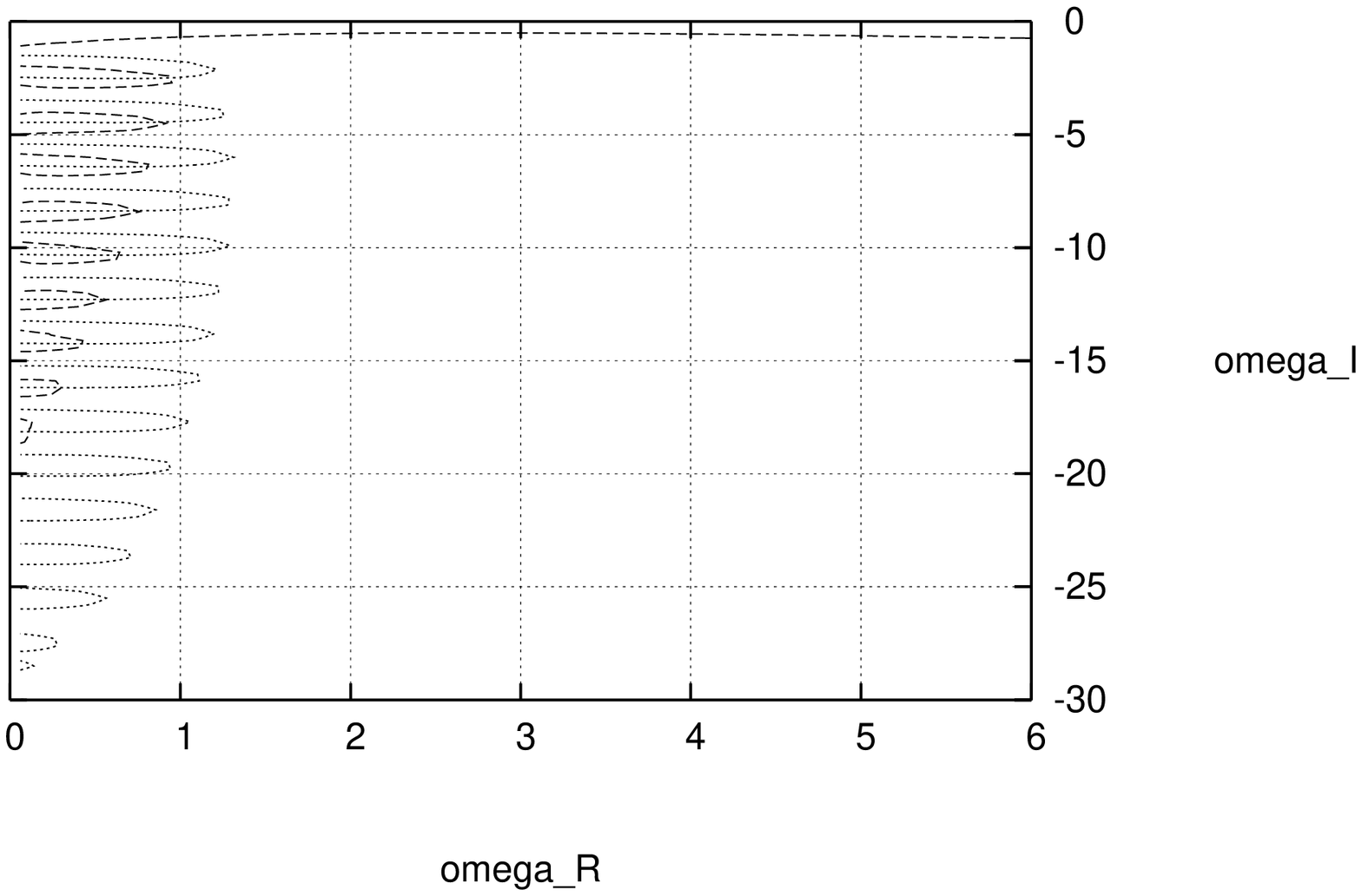}
\caption{Scalar QNMs in four dimensions at $r_+=0.99, \xi=1.0.$}
\label{d4_099_10}
\end{figure}
We start with the case $r_+=0.95$ (below the critical point),
$\xi=5.0,$ depicted in figure \ref{d4_095_50}. We observe that
there exists a finite number of QNMs and positive slope. Similar
results may be seen in figure \ref{d4_099_50}, corresponding to
$r_+=0.99$ and $\xi=5.0.$ In the latter case the real parts do not
change as much between successive modes. Going to smaller values
of $\xi$ the most important changes are that the region of
propagating modes becomes smaller, while the real parts take
smaller values. This is apparent upon inspection of figure
\ref{d4_099_30} (for $\xi=3.0$) and figure \ref{d4_099_10} (for
$\xi=1.0$). In the latter case purely dissipative modes appear.
There exist no modes at all for imaginary parts absolutely larger
than $35.$. The purely dissipative modes disappear for larger
values of $\xi;$ for $\xi=2$ it is difficult to detect them. This
fact is consistent with the analytic result that there exist
critical values of $\xi,$ above which only propagating modes
exist. Thus, for small enough values of $\xi$ we have several
propagating modes with relatively small absolute values of their
imaginary parts and positive slope, followed by a finite number of
purely dissipative modes. For larger values of $\xi$ only a finite
number of propagating modes exists.

To illustrate the features of the case with $r_+>1.0,$ (above the
critical point), we give in figure ~\ref{d4_110_10} the results
for $r_+=1.10, \ \xi=1.0.$ The slope is negative right from the
beginning and no purely dissipative modes emerge.

\begin{figure}[!ht]
\centering
\includegraphics[angle=-90,scale=0.5]{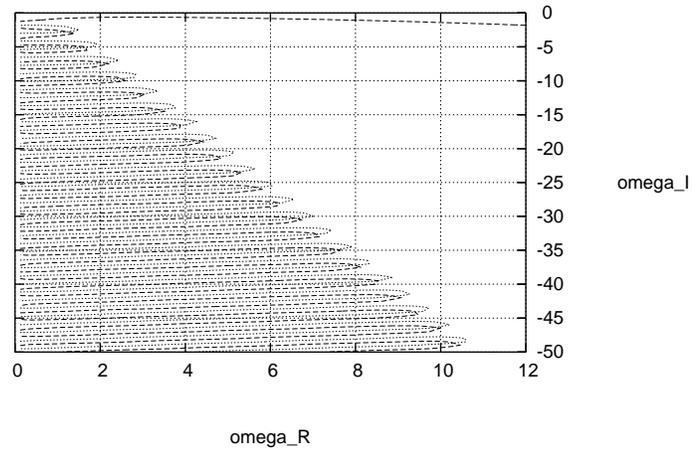}
\caption{Scalar QNMs in four dimensions at $r_+=1.10, \xi=1.0$}
\label{d4_110_10}
\end{figure}


Another set of results refers to the lowest modes, in particular
the behaviour of their real parts as functions of $\xi.$ The
analytical results derived from eq.~(\ref{eqar1}) along with the
numerical results are shown in figure \ref{realparts4}. The
agreement is good and it improves for large values of $\xi.$ In
particular, values of $\xi$ where these two real parts vanish
compare quite well with the analytically computed values.

\begin{figure}[!ht]
\centering
\includegraphics[angle=-90,scale=0.3]{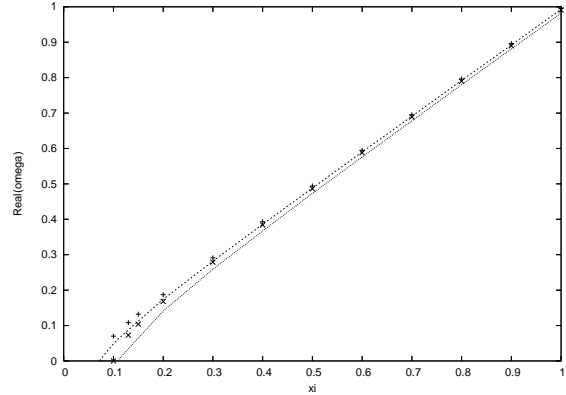}
\caption{The analytical results (eq.~(\ref{eqar1})) for the real
parts of the two lowest QNMs $\omega_0$ and $\omega_1$ vs $\xi$
for $d=4$ and $r_+=0.999$ are represented by lines, while their
numerical counterparts by points. The upper (lower) curve and the
corresponding points represent $\omega_0 \ (\omega_1).$}
\label{realparts4}
\end{figure}

In figure \ref{xicr4} we show the dependence of the critical
values $ \xi_0$ and $\xi_1$ on the horizon $r_+$. This figure
depicts the analytical results expressed by eq.~(\ref{eqana6a}),
as well as the corresponding numerical results. The agreement
between the two approaches is quite good.

\begin{figure}[!ht]
\centering
\includegraphics[angle=-90,scale=0.3]{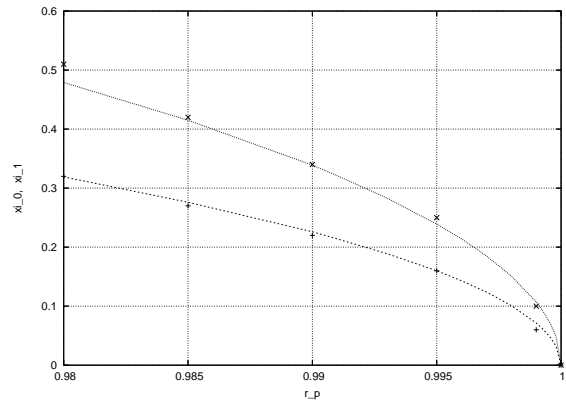}
\caption{The points represent numerical estimates of the critical
values $\xi_0$ and $\xi_1$ as functions of the horizon for $d=4,$
while lines give the corresponding analytical estimates
(eq.~(\ref{eqana6a})). The larger (lower) values represent $\xi_0
\ (\xi_1).$} \label{xicr4}
\end{figure}



\begin{table}[!ht]

\begin{center}
\begin{tabular}{|l|l|l|l|l|l|}

\hline

$\xi$ & $r_+$ & $\omega_0^{anal}$ & $\omega_0^{num}$ &
$\omega_1^{anal}$ & $\omega_1^{num}$ \\

\hline

$1.0$ & $0.95$ & $0.64-2.23i$ & $0.80 - 2.36i$ & $0.11 - 3.89i$
& $0.42 - 4.17i$ \\

\hline

$3.0$ & $0.95$ & $2.80-2.32i$ & $2.90 - 2.42i$ & $2.51 - 4.08i$
& $2.75 - 4.30i$ \\

\hline

$5.0$ & $0.95$ & $4.85-2.36i$ & $4.93 - 2.44i$ & $4.63 - 4.16i$
& $4.81 - 4.34i$ \\

\hline

$1.0$ & $0.99$ & $0.93-2.44i$ & $0.96 - 2.47i$ & $0.82 - 4.38i$
& $0.91 - 4.44i$ \\

\hline

$5.0$ & $0.99$ & $4.97-2.47i$ & $4.99 - 2.49i$ & $4.93 - 4.43i$
& $4.96 - 4.47i$ \\

\hline

$5.0$ & $0.999$ & $4.9971-2.4972i$ & $4.9985 - 2.4986i$ & $4.9926
- 4.4932i$ & $4.9963 - 4.4965i$ \\

\hline

$5.0$ & $0.9995$ & $4.9985-2.4986 i$ & $4.9993-2.4993 i$ &
$4.9963-4.4966 i$ & $4.9981-4.4982 i$ \\

\hline

$5.0$ & $0.9999$ & $4.9997-2.4997 i$ & $4.9998-2.4998 i$ &
$4.9993-4.4993 i$
& $4.9996-4.4996 i$ \\

\hline

\end{tabular}

\end{center}
\caption{Comparison of analytic {\em vs} numerical results for the
two lowest QNMs at $d=4$}\label{table1}
\end{table}

Agreement between analytical (eqs.~(\ref{eqana1}) and
(\ref{eqar1})) and numerical results for the two lowest QNMs is
further demonstrated in table \ref{table1}. Let us comment that
the two biggest critical values for $\xi$ in this case occur for
$r_+=0.95$ and read $\xi_{0}=0.50, \ \xi_{1}=0.76,$ well below the
values of $\xi$ considered here. The agreement is better as we
approach the value $r_+=1.0,$ as is evident in the last four rows
of table \ref{table1}.

Finally we depict in figure \ref{d4_lowest_1_2} the behaviour of
the lowest purely dissipative QNM as a function of $r_+$ for
$\xi=1.0$ and $\xi=2.0.$ It tends to $-\infty$ as the horizon
approaches the critical value $r_+=1.$

\begin{figure}[!t]
\centering
\includegraphics[angle=-90,scale=0.3]{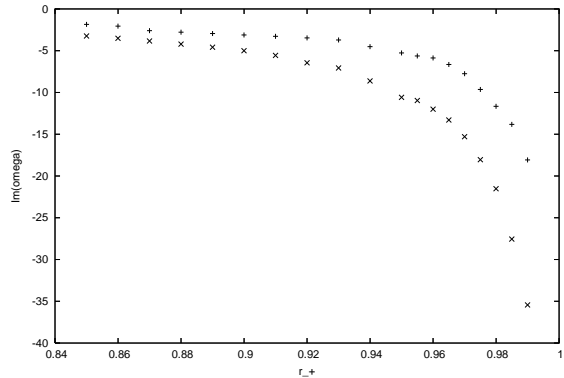}
\caption{Imaginary part of the lowest purely dissipative QNM in
four dimensions {\em vs} $r_+$ for $\xi=1.0$ (crosses) and
$\xi=2.0.$ } \label{d4_lowest_1_2}
\end{figure}

\subsection{$d=5$ and $d=6$}

Only quantitative changes appear in the five-dimensional case. We
depict sample results in figure \ref{d5_099_30}. Comparing against
figure \ref{d4_099_30}, let us only note the quantities
$|\Re(\omega)|$ change more between successive modes. We also note
that in five dimensions the propagating modes coexist with purely
dissipative modes in the same range of imaginary parts. Of course,
there are no propagating modes with sufficiently large values of
$|\Im(\omega)|,$ i.e., the propagating modes disappear and only
dissipative modes survive.

\begin{figure}[!ht]
\centering
\includegraphics[angle=-90,scale=0.5]{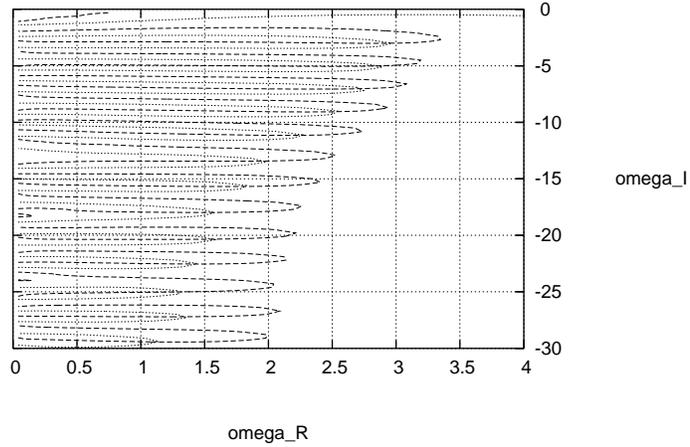}
\caption{Scalar QNMs in five dimensions at $r_+=0.99, \xi=3.0$}
\label{d5_099_30}
\end{figure}

As one moves to $d=6,$ numerical difficulties become more severe
and QNMs with too large imaginary parts become increasingly harder
to explore. We show sample results in figure
\ref{d6_rp=0.9999_10}. It seems that we encounter a qualitative
change. The propagating QNMs have positive slope in the region of
(absolutely) small imaginary parts, then there is a number of
purely dissipative modes, while for even larger imaginary parts a
new branch appears, which has negative slope.

\begin{figure}[!ht]
\centering
\includegraphics[angle=-90,scale=0.5]{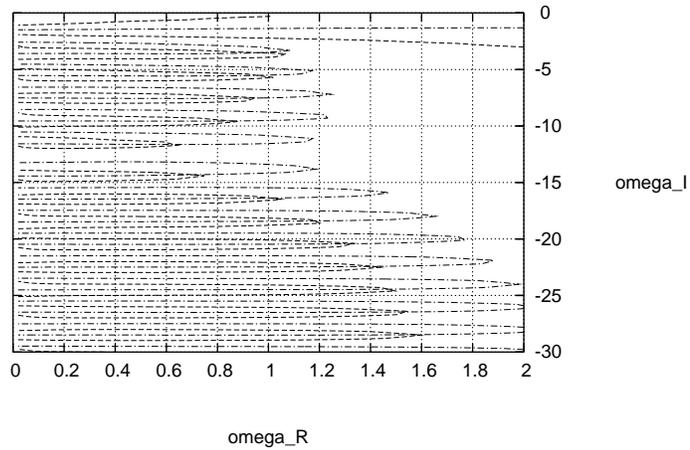}
\caption{Scalar QNMs in six dimensions at $r_+=0.9999, \xi=1.0.$}
\label{d6_rp=0.9999_10}
\end{figure}

We also study the real parts of the lowest modes for $d=5$ as
functions of $\xi$ and present both the analytical results derived
from eq.~(\ref{eqar1}) and their numerical counterparts in figure
\ref{realparts5}. The numerical results lie a bit above the
prediction and the agreement is not so good as in the $d=4$ case.
This feature is even more pronounced for $d=6.$ The results are
depicted in figure \ref{realparts6}. It is evident that the most
sizeable discrepancies occur for $\omega_1.$ In particular, the
numerical results appear more steep and the corresponding critical
value for $\xi_1$ is substantially  bigger than the analytical
prediction. These results show that first-order perturbation
theory is inadequate in this case - one needs to include
higher-order perturbative effects.

\begin{figure}[!ht]
\centering
\includegraphics[angle=-90,scale=0.3]{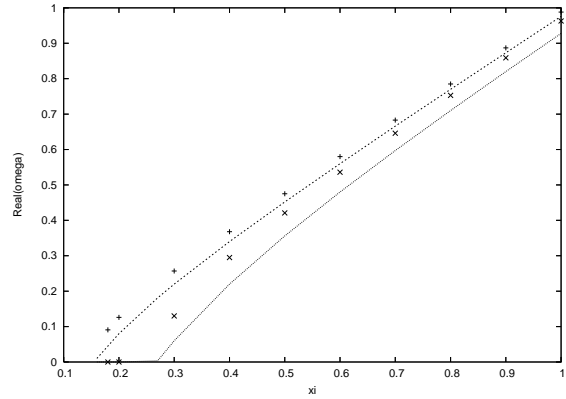}
\caption{The analytical results (eq.~(\ref{eqar1})) for the real
parts of the two lowest QNMs $\omega_0$ and $\omega_1$ vs $\xi$
for $d=5$ and $r_+=0.999$ are represented by lines, while their
numerical counterparts by points. The upper (lower) curve and the
corresponding points represent $\omega_0 \ (\omega_1).$}
\label{realparts5}
\end{figure}

\begin{figure}[!ht]
\centering
\includegraphics[angle=-90,scale=0.3]{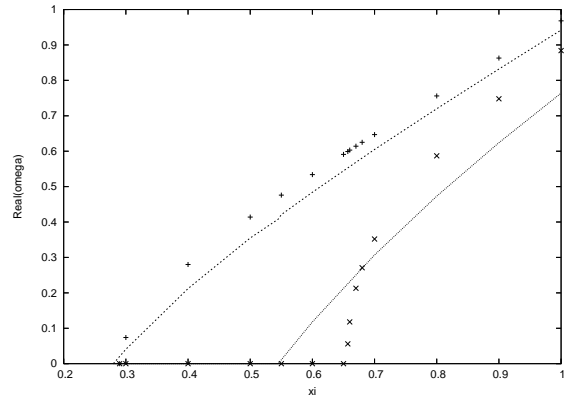}
\caption{The analytical results (eq.~(\ref{eqar1})) for the real
parts of the two lowest QNMs $\omega_0$ and $\omega_1$ vs $\xi$
for $d=6$ and $r_+=0.999$ are represented by lines, while their
numerical counterparts by points. The upper (lower) curve and the
corresponding points represent $\omega_0 \ (\omega_1).$}
\label{realparts6}
\end{figure}

In figure \ref{xicr5} we show the dependence of the critical
values $ \xi_0$ and $\xi_1$ on the horizon $r_+$ in five
dimensions. This figure depicts both the analytical results
expressed by equation (\ref{eqana6a}) and the corresponding
numerical results. Similar results are presented in figure
\ref{xicr6} for six dimensions. The agreement between analytical
and numerical results is satisfactory in all cases.


\begin{figure}[!t]
\centering
\includegraphics[angle=-90,scale=0.3]{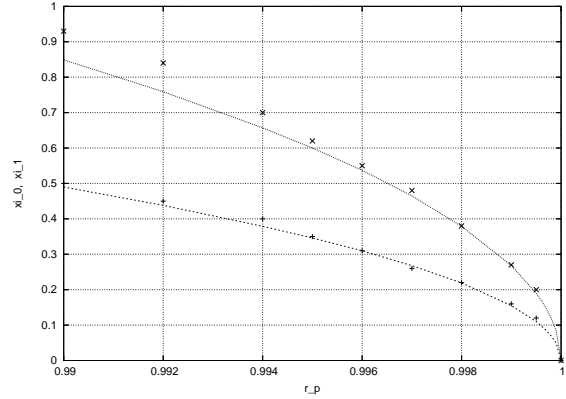}
\caption{The points represent numerical estimates of the critical
values $\xi_0$ and $\xi_1$ as functions of the horizon for $d=5,$
while lines give the corresponding analytical estimates
(eq.~(\ref{eqana6a})). The larger (lower) values represent $\xi_0
\ (\xi_1).$ } \label{xicr5}
\end{figure}
\begin{figure}[!t]
\centering
\includegraphics[angle=-90,scale=0.3]{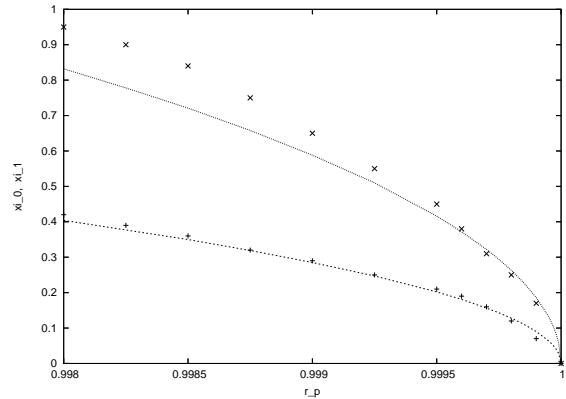}
\caption{The points represent numerical estimates of the critical
values $\xi_0$ and $\xi_1$ as functions of the horizon for $d=6,$
while lines give the corresponding analytical estimates
(eq.~(\ref{eqana6a})). The larger (lower) values represent $\xi_0
\ (\xi_1).$ } \label{xicr6}
\end{figure}

Table \ref{table2} contains the comparison of analytical versus
numerical results for the two lowest QNMs at $d=5.$ The agreement
is fairly good. Let us comment that the two critical values for
$\xi$ read in this case: $\xi_{0}=1.09, \ \xi_{1}=1.90$ for
$r_+=0.95$ and and smaller for the remaining values of $r_+.$ The
values of $\xi$ considered lie above the critical values.

\begin{table}[!ht]
\begin{center}



\begin{tabular}{|l|l|l|l|l|l|}

\hline

$\xi$ & $r_+$ & $\omega_0^{anal}$ & $\omega_0^{num}$ &
$\omega_1^{anal}$ & $\omega_1^{num}$ \\

\hline

$3.0$ & $0.95$ & $2.60-3.00i$ & $2.82 - 3.02i$ & $1.80 - 5.00i$
& $2.43 - 5.17i$ \\

\hline

$5.0$ & $0.95$ & $4.76-3.00i$ & $4.89 - 3.01i$ & $4.28 - 5.00i$
& $4.67 - 5.05i$ \\

\hline

$3.0$ & $0.99$ & $2.92-3.00i$ & $2.96 - 3.00i$ & $2.76 - 5.00i$
& $2.88 - 5.00i$ \\

\hline

$5.0$ & $0.99$ & $4.95-3.00i$ & $4.97 - 3.00i$ & $4.86 - 5.00i$
& $4.93 - 5.00i$ \\

\hline

$5.0$ & $0.999$ & $4.9952-3.0000i$ & $4.9976 - 3.0000i$ & $4.9856
- 5.0000i$ & $4.9928 - 5.0000i$ \\

\hline

$5.0$ & $0.9995$ & $4.9976-3.0000i$ & $4.9988 - 3.0000i$ & $4.9928
- 5.0000i$ & $4.9964 - 5.0000i$ \\

\hline

$5.0$ & $0.9999$ & $4.9995-3.0000i$ & $4.9998 - 3.0000i$ & $4.9986
- 5.0000i$ & $4.9993 - 5.0000i$ \\

\hline

\end{tabular}

\end{center}
\caption{Comparison of analytic {\em vs} numerical results for the
two lowest QNMs at $d=5$}\label{table2}
\end{table}

Table \ref{table3} contains the comparison of analytic versus
numerical results for the two lowest QNMs at $d=6.$ The agreement
is fairly good. Let us comment that the two critical values for
$\xi$ read: $\xi_{0}=0.90, \ \xi_{1}=1.86$ for $r_+=0.99.$ They
are smaller for the remaining values of $r_+.$ The values of $\xi$
considered lie above the critical values.

\begin{table}[!ht]

\begin{center}

\begin{tabular}{|l|l|l|l|l|l|}

\hline

$\xi$ & $r_+$ & $\omega_0^{anal}$ & $\omega_0^{num}$ &
$\omega_1^{anal}$ & $\omega_1^{num}$ \\

\hline

$3.0$ & $0.99$ & $2.89-3.58i$ & $2.91 - 3.65i$ & $2.59 - 5.78i$
& $3.06 - 5.77i$ \\

\hline

$5.0$ & $0.99$ & $4.95-3.54i$ & $4.81 - 3.61i$ & $4.82 - 5.64i$
& $4.90 - 5.97i$ \\

\hline

$5.0$ & $0.999$ & $4.9951-3.5042i$ & $4.9822 - 3.5140i$ & $4.9822
- 5.5141i$ & $5.0027 - 5.5393i$ \\

\hline

$5.0$ & $0.9995$ & $4.9976-3.5021i$ & $4.9911 - 3.5071i$ & $4.9911
- 5.5070i$ & $5.0015 - 5.5193i$ \\

\hline

$5.0$ & $0.9999$ & $4.9995-3.5004i$ & $4.9982 - 3.5014i$ & $4.9982
- 5.5014i$ & $5.0003 - 5.5038i$ \\

\hline

\end{tabular}

\end{center}
\caption{Comparison of analytic {\em vs} numerical results for the
two lowest QNMs at $d=6$}\label{table3}
\end{table}

In figure \ref{o_diss_5} we show exact values of the imaginary
part of the lowest purely dissipative mode for $\xi=1$ and $\xi=2$
versus $r_+$ for $d=5,$ along with a fit encoding the divergence
at the critical point $r_+=1.$ These results are successfully
reproduced numerically as can easily be seen in figure
\ref{d5_lowest_1_2}. The corresponding results for $d=6,$ where no
analytical results are available, are contained in figure
\ref{d6_lowest_1_2}.

\begin{figure}[!ht]
\centering
\includegraphics[scale=0.5]{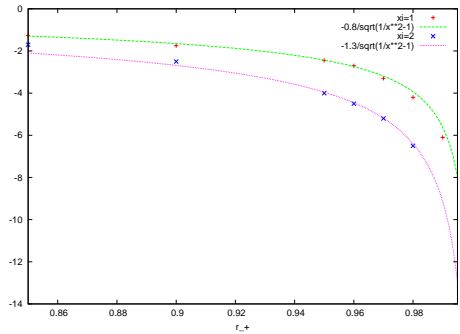}
\caption{Exact results for the imaginary part of the lowest purely
dissipative  mode for $\xi=1$ and $\xi=2$ {\em vs} $r_+$ for
$d=5.$} \label{o_diss_5}
\end{figure}
\begin{figure}[!ht]
\centering
\includegraphics[angle=-90,scale=0.3]{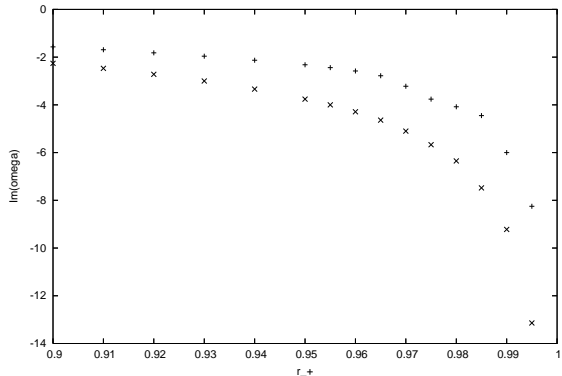}
\caption{Numerical results for the results for the imaginary part
of the lowest purely dissipative QNM in five dimensions {\em vs}
$r_+$ for $\xi=1.0$ (crosses) and $\xi=2.0.$}
\label{d5_lowest_1_2}
\end{figure}
\begin{figure}[!ht]
\centering
\includegraphics[angle=-90,scale=0.3]{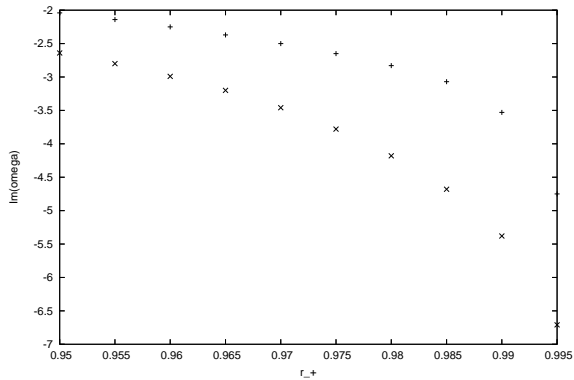}
\caption{Imaginary part of the lowest purely dissipative QNM in
six dimensions {\em vs} $r_+$ for $\xi=1.0$ (crosses) and
$\xi=2.0.$} \label{d6_lowest_1_2}
\end{figure}

\section{Conclusions}
\label{sec5}

We have studied the perturbative behaviour of the topological-AdS
black holes. We have calculated both analytically and numerically
the QNMs of scalar perturbations of these black holes.

Analytical calculations show that for small black holes  at any
dimension there is a critical point (at $r_+=1$) below which the
real part decreases with $n$, having a positive slope, whereas
above the critical point the oscillatory modes increase with a
negative slope. We also found that below the critical point there
is a critical value of $\xi$ below which there are purely decaying
modes while above the critical point there are only oscillatory
modes for any $\xi$. In five dimensions the QNMs of scalar
perturbations of TBH-AdS can be obtained explicitly from the Heun
function which solves the five-dimensional wave equation.

These results are also supported by  numerical investigations of
the QNMs. The numerical results show clearly a change of slope of
QNMs around a critical temperature for dimensions $d=4,5,6.$  For
larger dimensions the root finding algorithm is difficult to be
implemented. This is connected with the observation that for
higher dimensional theories the regularization needed for the
convergence of the series becomes less efficient.


\section*{Acknowledgments}

We thank E. Berti and V. Cardoso for their comments. Work
supported by the NTUA research program PEVE07. G.~S.~was supported
in part by the US Department of Energy under grant
DE-FG05-91ER40627.



\begin{thebibliography}{99}


\bibitem{KS}
  K.~D.~Kokkotas and B.~G.~Schmidt,
  Living Rev.\ Rel.\  {\bf 2}, 2 (1999)
  [arXiv:gr-qc/9909058].

\bibitem{N} H.-P. Nollert,
  Class.\ Quant.\ Grav.\  {\bf 16} (1999) R159.

\bibitem{CM}
  J.~S.~F.~Chan and R.~B.~Mann,
  Phys.\ Rev.\ D {\bf 55}, 7546 (1997)
  [arXiv:gr-qc/9612026];
  Phys.\ Rev.\ D {\bf 59}, 064025 (1999).

\bibitem{HH}
  G.~T.~Horowitz and V.~E.~Hubeny,
  Phys.\ Rev.\ D {\bf 62}, 024027 (2000)
  [arXiv:hep-th/9909056].

\bibitem{CL}
  V.~Cardoso and J.~P.~S.~Lemos,
  Phys.\ Rev.\ D {\bf 64}, 084017 (2001)
  [arXiv:gr-qc/0105103].



\bibitem{WLA}
  B.~Wang, C.~Y.~Lin and E.~Abdalla,
  Phys.\ Lett.\ B {\bf 481}, 79 (2000)
  [arXiv:hep-th/0003295].

\bibitem{kokkotas}
  E.~Berti and K.~D.~Kokkotas,
  Phys.\ Rev.\ D {\bf 67}, 064020 (2003)
  [arXiv:gr-qc/0301052].


\bibitem{Konoplya}
  R.~A.~Konoplya,
  Phys.\ Rev.\  D {\bf 66}, 044009 (2002)
  [arXiv:hep-th/0205142];
  V.~Cardoso, R.~Konoplya and J.~P.~S.~Lemos,
  Phys.\ Rev.\  D {\bf 68}, 044024 (2003)
  [arXiv:gr-qc/0305037].

 \bibitem{RHIC}
  I.~Arsene {\it et al.}  [BRAHMS Collaboration],
  Nucl.\ Phys.\  A {\bf 757}, 1 (2005)
  [arXiv:nucl-ex/0410020];
  K.~Adcox {\it et al.}  [PHENIX Collaboration],
  Nucl.\ Phys.\  A {\bf 757}, 184 (2005)
  [arXiv:nucl-ex/0410003];
  J.~Adams {\it et al.}  [STAR Collaboration],
  Nucl.\ Phys.\  A {\bf 757}, 102 (2005)
  [arXiv:nucl-ex/0501009].

\bibitem{Friess:2006kw}
  J.~J.~Friess, S.~S.~Gubser, G.~Michalogiorgakis and S.~S.~Pufu,
  JHEP {\bf 0704}, 080 (2007)
  [arXiv:hep-th/0611005].

  \bibitem{condensed_Matter}
  S.~A.~Hartnoll and P.~Kovtun,
  Phys.\ Rev.\  D {\bf 76}, 066001 (2007)
  [arXiv:0704.1160 [hep-th]];
  S.~A.~Hartnoll, P.~K.~Kovtun, M.~Muller and S.~Sachdev,
  Phys.\ Rev.\  B {\bf 76}, 144502 (2007)
  [arXiv:0706.3215 [cond-mat.str-el]];
  S.~A.~Hartnoll and C.~P.~Herzog,
  Phys.\ Rev.\  D {\bf 76}, 106012 (2007)
  [arXiv:0706.3228 [hep-th]];
  S.~S.~Gubser,
  arXiv:0801.2977 [hep-th].




\bibitem{Hartnoll:2008vx}
  S.~A.~Hartnoll, C.~P.~Herzog and G.~T.~Horowitz,
  Phys.\ Rev.\ Lett.\  {\bf 101}, 031601 (2008)
  [arXiv:0803.3295 [hep-th]];
  S.~A.~Hartnoll, C.~P.~Herzog and G.~T.~Horowitz,
  arXiv:0810.1563 [hep-th].


\bibitem{Cardoso:2004up}
  V.~Cardoso, J.~Natario and R.~Schiappa,
  J.\ Math.\ Phys.\  {\bf 45}, 4698 (2004)
  [arXiv:hep-th/0403132].

 \bibitem{cft}
  J.~M.~Maldacena,
  Adv.\ Theor.\ Math.\ Phys.\  {\bf 2}, 231 (1998)
  [Int.\ J.\ Theor.\ Phys.\  {\bf 38}, 1113 (1999)]
  [arXiv:hep-th/9711200];
  E.~Witten,
  Adv.\ Theor.\ Math.\ Phys.\  {\bf 2}, 253 (1998)
  [arXiv:hep-th/9802150];
  S.~S.~Gubser, I.~R.~Klebanov and A.~M.~Polyakov,
  Phys.\ Lett.\  B {\bf 428}, 105 (1998)
  [arXiv:hep-th/9802109].




\bibitem{phase1} S. Hawking and D. Page, Commun. Math. Phys. 87 (1983) 577.

\bibitem{phase2}
  E.~Witten,
  Adv.\ Theor.\ Math.\ Phys.\  {\bf 2}, 505 (1998)
  [arXiv:hep-th/9803131].

\bibitem{Koutsoumbas:2006xj}
  G.~Koutsoumbas, S.~Musiri, E.~Papantonopoulos and G.~Siopsis,
  JHEP {\bf 0610}, 006 (2006)
  [arXiv:hep-th/0606096].


\bibitem{Koutsoumbas:2008pw}
  G.~Koutsoumbas, E.~Papantonopoulos and G.~Siopsis,
  JHEP {\bf 0805}, 107 (2008)
  [arXiv:0801.4921 [hep-th]].

\bibitem{Reall:2001ag}
  H.~S.~Reall,
  Phys.\ Rev.\  D {\bf 64}, 044005 (2001)
  [arXiv:hep-th/0104071].




\bibitem{Jing:2008an}
  J.~Jing and Q.~Pan,
  Phys.\ Lett.\  B {\bf 660}, 13 (2008)
  [arXiv:0802.0043 [gr-qc]].

\bibitem{Berti:2008xu}
  E.~Berti and V.~Cardoso,
  Phys.\ Rev.\  D {\bf 77}, 087501 (2008)
  [arXiv:0802.1889 [hep-th]].

\bibitem{He:2008im}
  X.~He, B.~Wang, S.~Chen, R.~G.~Cai and C.~Y.~Lin,
  arXiv:0802.2449 [hep-th].

\bibitem{Shen:2007xk}
  J.~Shen, B.~Wang, C.~Y.~Lin, R.~G.~Cai and R.~K.~Su,
  JHEP {\bf 0707}, 037 (2007)
  [arXiv:hep-th/0703102];
  X.~Rao, B.~Wang and G.~Yang,
  Phys.\ Lett.\  B {\bf 649}, 472 (2007)
  [arXiv:0712.0645 [gr-qc]].

 \bibitem{Lemos}
  J.~P.~S.~Lemos and V.~T.~Zanchin,
  Phys.\ Rev.\  D {\bf 54}, 3840 (1996)
  [arXiv:hep-th/9511188];
  J.~P.~S.~Lemos,
  Phys.\ Lett.\  B {\bf 353}, 46 (1995)
  [arXiv:gr-qc/9404041].

  \bibitem{mann}
  R.~B.~Mann,
  Class.\ Quant.\ Grav.\  {\bf 14}, L109 (1997)
  [arXiv:gr-qc/9607071];
  R.~B.~Mann,
  Nucl.\ Phys.\ B {\bf 516}, 357 (1998)
  [arXiv:hep-th/9705223].

\bibitem{Vanzo:1997gw}
  L.~Vanzo,
  Phys.\ Rev.\ D {\bf 56}, 6475 (1997)
  [arXiv:gr-qc/9705004].


\bibitem{Brill:1997mf}
  D.~R.~Brill, J.~Louko and P.~Peldan,
  Phys.\ Rev.\ D {\bf 56}, 3600 (1997)
  [arXiv:gr-qc/9705012].



\bibitem{Birmingham}
  D.~Birmingham,
  Class.\ Quant.\ Grav.\  {\bf 16}, 1197 (1999)
  [arXiv:hep-th/9808032].

\bibitem{Wang:2001tk}
  B.~Wang, E.~Abdalla and R.~B.~Mann,
  Phys.\ Rev.\  D {\bf 65}, 084006 (2002)
  [arXiv:hep-th/0107243].

\bibitem{Aros:2002te}
  R.~Aros, C.~Martinez, R.~Troncoso and J.~Zanelli,
  Phys.\ Rev.\  D {\bf 67}, 044014 (2003)
  [arXiv:hep-th/0211024].

\bibitem{Birmingham:2006zx}
  D.~Birmingham and S.~Mokhtari,
  Phys.\ Rev.\  D {\bf 74}, 084026 (2006)
  [arXiv:hep-th/0609028].

\bibitem{Sheykhi:2007wg}
  A.~Sheykhi,
  arXiv:0709.3619 [hep-th];
  M.~Nadalini, L.~Vanzo and S.~Zerbini,
  arXiv:0710.2474 [hep-th].

\bibitem{Myung:2006tg}
  Y.~S.~Myung,
  Phys.\ Lett.\  B {\bf 645}, 369 (2007)
  [arXiv:hep-th/0603200];
  arXiv:0801.2434 [hep-th].


\bibitem{Gibbons:2002pq}
  G.~Gibbons and S.~A.~Hartnoll,
  Phys.\ Rev.\  D {\bf 66}, 064024 (2002)
  [arXiv:hep-th/0206202].

\bibitem{Birmingham:2007yv}
  D.~Birmingham and S.~Mokhtari,
  Phys.\ Rev.\  D {\bf 76}, 124039 (2007)
  [arXiv:0709.2388 [hep-th]].

\bibitem{Emparan:1999gf}
  R.~Emparan,
  JHEP {\bf 9906}, 036 (1999)
  [arXiv:hep-th/9906040].































\end{thebibliography}
\end{document}